\documentclass[aps,twocolumn,aps,floatfix]{revtex4-1}
\usepackage{graphicx}
\usepackage{color}
\usepackage{amsmath}
\usepackage{fullpage}
\usepackage{hyperref}
\usepackage{mathrsfs}
\usepackage{placeins}

\graphicspath{ {C:/Users/owner/Documents/Itamar/GDrive/Papers/Paper1/fig/} }

\hypersetup{
    colorlinks=true,       
    linkcolor=black, 
    citecolor=blue, 
    filecolor=magenta,      
    urlcolor=cyan           
}

\begin{document}

\title{Electrically controlled mutual interactions of flying waveguide dipolaritons}
\author{Itamar Rosenberg}
\author{Yotam Mazuz-Harpaz}
\author{Ronen Rapaport}
\affiliation{Racah Institute of Physics, The Hebrew University of Jerusalem, Jerusalem 91904, Israel}
\author{Kenneth West}
\author{Loren Pfeiffer}
\affiliation{Department of Electrical Engineering, Princeton University, Princeton, New Jersey 08544, USA}

\begin{abstract}
We show that with a system of electrically-gated wide quantum wells embedded inside a simple dielectric waveguide structure, it is possible to excite, control, and observe waveguided exciton polaritons that carry an electric dipole moment. We demonstrate that the energy of the propagating dipolariton can be easily tuned using local electrical gates, that their excitation and extraction can be easily done using simple evaporated metal gratings, and that the dipolar interactions between polaritons and between polaritons and excitons can also be controlled by the applied electric fields. This system of gated flying dipolaritons thus exhibit the ability to locally control both the single polariton properties as well as the interactions between polaritons, which should open up opportunities for constructing complex polaritonic circuits and for studying strongly-interacting, correlated polariton gases.   
\end{abstract}
\maketitle
\section{Introduction}
Exciton polaritons are the dressed states of quantum well (QW) excitons which are strongly coupled to confined photons. Following their first experimental discovery\cite{weisbuch_observation_1992}, 
exciton-polaritons with photons confined in a cavity mode of a semiconductor microcavity (MC) have displayed a range of quantum collective phenomena \cite{imamoglu_quantum_1996,dang_stimulation_1998,senellart_nonlinear_1999,kasprzak_bose-einstein_2006,balili_bose-einstein_2007,lagoudakis_quantized_2008,roumpos_single_2011,utsunomiya_observation_2008,amo_collective_2009} previously observed only in cold atomic gases. Exciton-polaritons are also promising for realizations of new polariton based logic circuitry which, if successful, would enable extremely fast transfer and processing of coherent information packets\cite{menon_nonlinear_2010}. Over the last few years, various building blocks for polariton-based circuits where demonstrated \cite{amo_excitonpolariton_2010,ferrier_interactions_2011,cristofolini_optical_2013, gao_polariton_2012,ballarini_all-optical_2013,sturm_all-optical_2014,nguyen_realization_2013}. In spite of the continuously growing number of experimental demonstrations of different device concepts, MC-polaritons have several significant drawbacks when considering complex large scale polaritonic circuits:
Due to the inherently thick geometry of the MC, the polaritons reside very far from the surface of the chip. As a result, control over the polariton confinement potential is possible only after complex  processing of the distributed Bragg reflectors (DBRs) \cite{zhang_zero-dimensional_2014}. This large DBR thickness also strongly limits the ability to locally shape the polariton potential landscape using spatially varying electrical fields applied via small electric gates, as the gates must be placed very far from the QWs plane. Moreover, the polaritons lifetime and thus their maximal propagation length are inherently limited by the finite reflectivity of the DBRs. MC-polaritons also leak as they propagate, thus the position of their out-coupling is hard to control. In addition, the QW excitons are neutral, and their interactions are therefore short ranged\cite{ciuti_role_1998} and cannot be tuned (The typical interaction strength in GaAs QWs is of the order of $5$ $\mu eV\mu m^2$\cite{tassone_exciton-exciton_1999}), limiting the tunability of nonlinearities required in polaritonic devices, and the ability to use them for studying strongly correlated systems \cite{laikhtman_exciton_2009,shilo_particle_2013}.
Recently, it was demonstrated that partially indirect MC-polaritons can form from an admixture of direct and indirect excitons in an electrically biased asymmetric double QW structure \cite{christmann_oriented_2011}. It was calculated \cite{byrnes_effective_2014} that these polaritons should possess electric dipole moments due to the e-h charge separation of their indirect excitonic part (thus they were termed dipolaritons) \cite{cristofolini_coupling_2012}. However a direct signature of their dipolar nature, exhibited as dipole-dipole interactions between polaritons has not yet been observed experimentally, and observing and controlling dipolar interactions between polaritons is still an open challenge.

In this work we demonstrate that by integrating electrically-gated wide QWs into simple photonic waveguide (WG) geometry, we can form dipolar WG-polaritons, allowing better control and additional functionality for the realization of future polaritonic devices. We specifically demonstrate that the energy of the polaritons can be widely tuned locally via the application of small voltages on lithographically defined evaporated gates\footnote{We note that electrical tuning of polaritons in MCs was first demonstrated by Fisher et. al. \cite{fisher_electric-field_1995}}. Furthermore we show that the applied voltage induces a highly tunable electric-dipole to the excitonic part of the polariton. This induced dipole leads to an observed repulsive dipole-dipole interactions between the dipolaritons. We also show that a controlled out-coupling of the polariton signal is easily achieved by properly designed metallic out-couplers. 

\begin{figure}[t]
  \centering
  \includegraphics[width=0.5\textwidth]{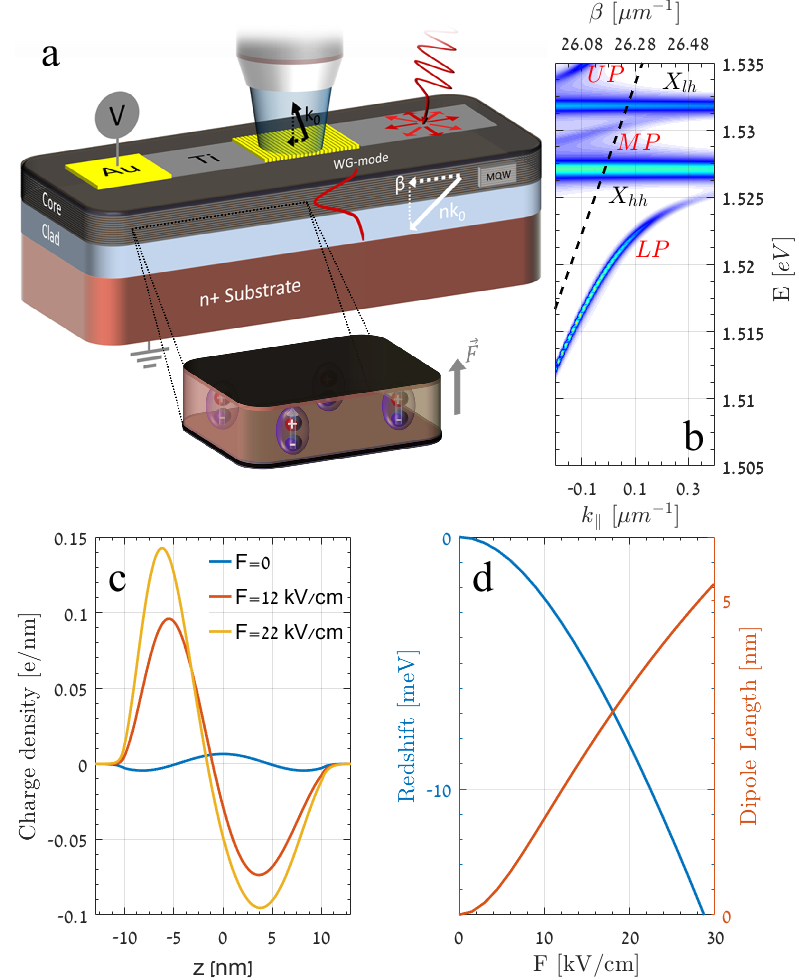}
  \caption{(a) An Illustration of the experimental scheme. A non resonant excitation through a thin Ti electrode is used to excite polaritons which propagate with a propagation constant $\beta$. A gold grating coupler with a periodicity of 240 nm is used to read out the signal. (b) Numerical calculation of the expected dispersion from our device in the absence of an electric field. The bare heavy hole ($X_{hh}$) and light hole ($X_{lh}$) excitons are marked as well as The LP, MP and UP polariton branches. The bare WG mode is illustrated by the dashed line. c) Calculations of the charge distribution along the growth direction for $F=0$, $12$ and $22 kV/cm$. (The non-uniform distribution when $\mathbf{F}=0$ originates from the mass difference between the electron and the hole which makes the electron wave function more extended.)  (d) Calculations of the exciton red shift and effective dipole length as a function of $\mathbf{F}$.}
    \label{fig:fig1}
\end{figure}
The general structure and physical concept are plotted in Fig.\ref{fig:fig1}(a), where a set of 12, 20 nm wide, Al$_{0.4}$Ga$_{0.6}$As/GaAs/Al$_{0.4}$Ga$_{0.6}$As QWs serve as the core of a slab waveguide on top of a 500 nm bottom clad layer of Al$_{0.8}$Ga$_{0.2}$As, and with no top clad layer. The details of the sample structure and the experimental setup are given in Appenix \ref{App:exp_details}. 
The WG polaritons result from the strong coupling of the propagating waveguide TE-mode of the structure \footnote{The heavy and light-hole excitons couple strongly also to the TM waveguide modes but with a weaker coupling strength. We do not address TM-polaritons in this paper, see Appendix \ref{App:TE_TM} for more information.}, having a WG propagation constant $\beta$, and the heavy-hole and light-hole excitonic optical transitions of the QWs having an in-plane $k$-vector $k_{\parallel}^X=\beta$. This strong coupling leads to three polariton branches, the Upper (UP), middle (MP) and lower (LP) branches, that exhibit an anti-crossing in their dispersion around $\beta_0$ that satisfy  $E_X(\beta_0)=E_{p}(\beta_0)$ i.e., the crossing points of the bare modes dispersion, as was shown in Refs. \cite{bajoni_exciton_2009,walker_exciton_2013,walker_ultra-low-power_2015}. The calculated dispersion of the system, $E(\beta)$, based on a transfer matrix formalism (see Appendix \ref{App:design_simulations}) is plotted in Fig.\ref{fig:fig1}(b). Since here the optical mode is confined by total internal reflection rather than by reflection from mirrors,  the resultant WG polaritons can propagate with much less optical losses, and at velocities determined by $\beta$, which are an order of magnitude larger than those typical for  MC-polaritons\cite{walker_exciton_2013}.
On top of the sample, a 10 nm thick, semitransparent Ti channel was deposited. This channel serves as a top electrode, to which voltage can be applied with respect to the $n^+$ back gate, resulting in a constant electric field ($\mathbf{F}$) directed perpendicular to the QWs plane (see inset of Fig.\ref{fig:fig1}(a)). The effect of the electric field is a quadratic Stark energy red-shift of the exciton resonance \cite{bastard_variational_1983}, leading to a red shift of the energy of all the polaritonic branches at the anti-crossing point and to a decrease of $\beta_0$. Furthermore, The applied field induces a net charge separation along the z-direction, inducing a net electric dipole moment for the exciton and thus for the WG-polaritons. Both the energy of the WG-polaritons and their dipole moment are controlled by the applied field. Fig.\ref{fig:fig1}(c,d) show self consistent Schr\"odinger-Poisson solutions of the charge density, the exciton energy shift and the effective dipole of the exciton (Due to the overlap between the wavefunctions of the holes and the electrons, the effective dipole length which controls the dipolar interactions, is smaller the the actual charge separation. See Appendix \ref{App:dipole} for more information) for several values of ($\mathbf{F}$) respectively. 
\section{Electric field control of the single polariton properties}
We first focus on measurements of the single WG-polariton properties and their dependence on $\mathbf{F}$. To excite these WG-polaritons, a non-resonant CW laser (at 780 nm) is focused on one end of the gated channel, which creates a reservoir of uncoupled excitons at the excitation spot. A large fraction of the excitons relax and accumulate at the bottleneck of the LP branch of the polariton dispersion, just below the anti-crossing point\cite{tassone_exciton-exciton_1999}. These bottleneck WG-polaritons reside around $\beta\sim 26.2\mu m^{-1}\gg 0$ where the dispersion is steep and thus they have a very high group velocity of $v_g=\frac{1}{\hbar}\frac{dE}{d\beta}\sim 67\mu m/ps$. In this sense the WG-polaritons are naturally "flying". Some of these WG-polaritons propagate along the Ti channel, so to extract them out of the WG at a specific location, we deposit an Au grating coupler on top of the Ti channel. The periodicity of the grating is chosen so that the magnitude of its reciprocal lattice vector matches the parallel momentum of the anti-crossing point of the polariton dispersion. As a result, the polariton dispersion around the anti-crossing point shifts towards zero parallel momentum and is coupled out by the grating, where their photoluminescence (PL) can be measured. Fig.\ref{fig:fig1}(b) shows the calculated dispersion of the emitted light after it is coupled out by the metal grating, as a function of $k_{||}$ (defined in Fig.\ref{fig:fig1}(a), see calculation details in Appendix \ref{App:design_simulations}).
\begin{figure}[h]
  \centering
  \includegraphics[width=0.5\textwidth]{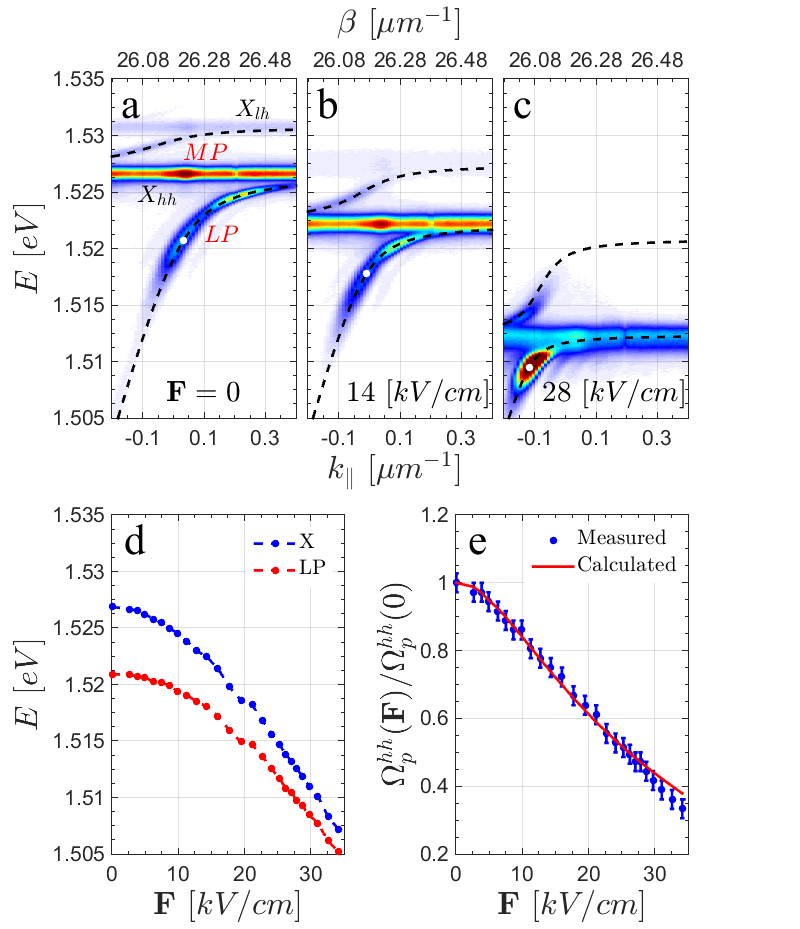}
  \caption{(a-c) The measured dispersion at three different values of $\mathbf{F}$ (measured at 30 K), fitted with a coupled oscillator model (dashed). A red shift of the entire polaritonic dispersion as $\mathbf{F}$ increases can be clearly seen. (d) The energies of the LP (below the avoided crossing point, marked by the white dots in (a-c)) and of the $X_{hh}$ as a function of $\mathbf{F}$. (e) The measured (blue dots) and calculated (red line) Rabi frequency ratio, $\Omega^{hh}_p(F)/\Omega^{hh}_p(0)$, as function of $\mathbf{F}$.}
    \label{fig:fig3}
\end{figure}
The Hamiltonian of the system neglecting particle interactions is given by:  
\begin{equation}\label{Coupled_Osc}
 \mathscr{H}_\beta=   \hbar\left( \begin{array}{ccc}
\omega_p(\beta) & \Omega^{hh}_{p}(\beta,\mathbf{F}) & \Omega^{lh}_{p}(\beta,\mathbf{F}) \\
\Omega^{hh}_{p}(\beta,\mathbf{F}) & \omega_{hh}(\beta,\mathbf{F}) & 0 \\
\Omega^{lh}_{p}(\beta,\mathbf{F}) & 0 & \omega_{lh}(\beta,\mathbf{F}) \end{array} \right)
  \end{equation}
Where $\omega_{hh}$, $\omega_{lh}$ and $\omega_{p}$ are the $X_{hh}$, $X_{lh}$, and photon bare eigenfrequencies respectively and $\Omega^{hh}_{p}$/$\Omega^{lh}_{p}$ are the corresponding Rabi frequencies. Two terms in Eq.\ref{Coupled_Osc}, depend on $\mathbf{F}$: the first is the bare exciton energy, which undergoes a quadratic Stark red shift due to the interaction with the electric field \cite{bastard_variational_1983}, as was depicted in Fig.\ref{fig:fig1}(d). The second is the Rabi frequency, which is proportional to the overlap integral between the envelope functions of the hole and the electron in the QWs. The charge separation induced by the perpendicular electric field (Fig.\ref{fig:fig1}(c)) reduces this integral,thus reducing the Rabi frequencies with increasing field. 
The effect of $\mathbf{F}$ on the measured dispersion can be clearly seen in Fig.\ref{fig:fig3}(a-c), where we show the PL dispersion coming out of the out-coupler for three different values of the applied electric field, and at a fixed excitation CW power of 37 $\mu W$. Here the excitation spot is close to the out-coupler and the temperature is relatively high (T=30 K), so all coupled and bare-exciton modes can be observed including the $X_{lh}$ and the MP branch. First, the red shift of the bare $X_{hh}$ energy with $\mathbf{F}$ results in a red shift of the whole polaritonic dispersion, and in a shift of the LP bottleneck to lower $\beta$ values. \textit{Due to the thin design of the sample}, a small voltage of only 4 V leads to large polariton energy shifts of 15 meV, as can be seen in Fig.\ref{fig:fig3}(d), and to a significant shift of $\beta$ at the bottleneck \footnote{The peculiar feature around $\mathbf{F}$=20 $kV/cm$ is still not well understood, but we note that it matches a kink in the IV curve which was measured during the experiment, and thus can be related to a resonant tunneling condition of the optically excited carriers}. This demonstrates the electric-field control capabilities of the WG-polariton system, which is a significant advantage over MC-polaritons.  
The other effect of $\mathbf{F}$ is the reduction of the MP-LP energy splitting with increasing $\mathbf{F}$ due to the reduction of the electron-hole wavefunction overlap. Fig.\ref{fig:fig3}(e) plots the relative reduction in $\Omega^{hh}_p$ with increasing $\mathbf{F}$. The experimental points of $\Omega^{hh}_p(F)/\Omega^{hh}_p(0)$ in Fig.\ref{fig:fig3}(e) were extracted by fitting the dispersion to Eq.\ref{Coupled_Osc}, shown as dotted black lines in Fig.\ref{fig:fig3}(a-c). The red line in Fig.\ref{fig:fig3}(e) is the theoretically calculated 
$\Omega^{hh}_p(F)/\Omega^{hh}_p(0)$ from the e-h wavefunction overlap using the numerical Schr\"odinger-Poisson solver. A very good agreement between experiment and theory is found. Even at a very significant LP energy shift of 10 meV, the MP-LP splitting is reduced by only less than a factor of 2, and the system is still well within the strong-coupling regime.
\begin{figure}[h]
  \centering
  \includegraphics[width=0.5\textwidth]{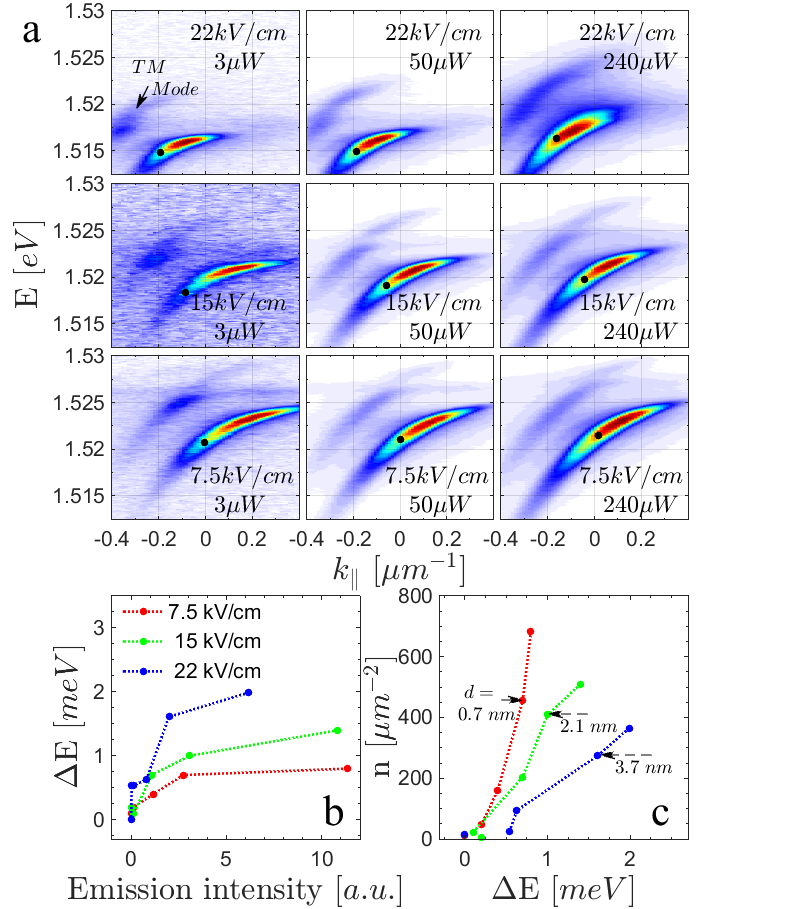}
  \caption{(a) Measured dispersion at different excitation powers and different values of  $\mathbf{F}$ (measured at 5 K). Each row represents a constant $\mathbf{F}$ while each column represents a constant excitation power. The increased energy blueshift ($\Delta E$) with increasing power is clearly seen, and is more significant at larger $\mathbf{F}$ values. The TM mode is marked by an arrow in the top left panel. (b) The  energy blueshift, $\Delta E$, measured where $\chi_x=\frac{1}{2}$ (marked by the black points on (a)) as a function of the LP emission intensity, for 3 different values of $\mathbf{F}$. (c) An estimation for the density of the dipoles extracted from the $\Delta E$ values in (b) using the mean field model. The value of the effective dipole for a selected point on each of the three curves is marked.}
    \label{fig:fig4}
\end{figure}
\section{Electric field control of induced dipolar interactions between polaritons}
Next we turn to observe the particle-particle interactions.
In the presence of an electric field the excitons, and thus the WG-polaritons become dipolar, with their dipole moments all pointing at the same direction, i.e., perpendicular to the QW plane. In this case the dipole-dipole interactions are repulsive, and thus are expected to yield an increase in the energy of the dipoles, which should result in a blue-shifted emission spectrum. Importantly, the interaction-induced energy blue-shift, $\Delta E$, is proportional to the dipole size $d$ \cite{laikhtman_exciton_2009,shilo_particle_2013}. $d$ in turn depends on the magnitude of $\mathbf{F}$ (as was shown in Fig.\ref{fig:fig1}(d)), and on the dipoles density, which due to electric screening reduces the effective Stark shift and thus tends to reduce the effective dipole. In absence of spatial correlations, the interaction energy of a population of dipoles with an average density $n$ is given by \cite{laikhtman_exciton_2009}: 
\begin{equation}\label{Eq:Capacitor_Formula}
    \Delta E=g(\mathbf{F},n)n =\frac{4\pi e^2 d(\mathbf{F},n)}{\epsilon}n
  \end{equation}
where $e$ is the electron charge, $\epsilon$ is the dielectric constant, and $d$ is the effective dipole length. We thus expect that a clear signature of dipolar interactions between WG-polaritons is an observation of a positive $\Delta E(F,n)$ which should increase with increasing dipole density $n$, and that its slope with respect to $n$ should increase with increasing $\mathbf{F}$. We also expect that for a fixed $\mathbf{F}$, $\Delta E$ should increase sub-linearly with $n$, since as mentioned above, $d$ decreases with increasing $n$. To check these predictions, we conducted a sequence of measurements  with a pulsed laser excitation (at $775 nm$ with $300 ps$ pulse duration) on a similar device as was described above at T=5 K. To observe only polariton-polariton interactions, the optical excitation was focused 15 $\mu m$ away from the out-coupler, and the emission from the out-coupler was measured using a gated intensified camera (PI-MAX) with a $10 ns$ exposure window overlapping the laser pulse. Since the bare $X_{hh}$ is too slow to reach the out-coupler within the short accumulation time of the camera,  we observed mostly emission from the fast, low-loss LPs, and the emission from the bare $X_{hh}$ was very weak, as can be seen from the measured emission spectra in Fig.\ref{fig:fig4}(a). Since the particles density increases with increasing excitation power, we expect a corresponding increase in the polariton energy. We conducted such power sequences with different fixed $\mathbf{F}$ values. In Fig.\ref{fig:fig4}(a) we present results from three of these sequences for 3 different values of $\mathbf{F}$. These measurements clearly show a blue shift of the LP line with increasing excitation power, which is indeed more significant as $\mathbf{F}$ is increased. We note that during these measurements no polarizer was used for the PL and thus the TM-mode was also measured (marked in the top left panel of Fig.\ref{fig:fig4}(a)). Since the LP density at the out-coupler is proportional to the LP emission intensity (the lifetime of the polaritons under the coupler is given by the coupling efficiency and is fixed for all experiments), we plot in Fig.\ref{fig:fig4}(b) the extracted $\Delta E$ of the LP-mode (below the avoided crossing point where the $X_{hh}$ fraction is $\frac{1}{2}$) as a function of the LP intensity. Again, a clear increase of $\Delta E$ with increasing intensity and thus with increasing polariton density is observed, indicating that this is indeed a many-body interaction effect. Furthermore, the increase is sublinear with the intensity, and its slope significantly increases with increasing $\mathbf{F}$, as is expected for dipolar interactions. The maximal measured $\Delta E$ increases from  0.8 meV at the lowest $\mathbf{F}$ value to 2 meV at the highest. On the contrary, such a dependence is not expected for non-dipolar, short-range interactions.

Using the Schr\"odinger-Poisson solver, we can extract the effective value of $d$ for each measurement point in Fig.\ref{fig:fig4}(b) and utilize it to calculate the $\mathbf{F}$-dependent coupling constant $g$ in Eq.\ref{Eq:Capacitor_Formula}. This yields $g\sim 5.7$ $\mu eV\mu m^2$ for $F=22$ $kV/cm$, $g\sim 3.6$ $\mu eV \mu m^2$ for $F=15$ $kV/cm$ and $g\sim 1.4$ $\mu eV \mu m^2$ for $F=7.5$ $kV/cm$\footnote{This should be compared with Ref.\cite{ferrier_interactions_2011} that reports coupling constant of $g\sim 9$ $\mu eV \mu m^2$ for polariton condensates in 10 $\mu m$ micropillar}. These values demonstrate that the dipolar interaction strength can be electrically controlled over a large tuning range, as was theoretically calculated in Ref. \cite{nalitov_voltage_2014}, and seems to be much larger than the prediction of Ref. \cite{byrnes_effective_2014}. Next, using these values of $g$ and the corresponding values of $\Delta E$, we estimated the local density of the dipoles $n$ at the out-coupler point, which we plot in Fig.\ref{fig:fig4}(c). These dipole densities are given by: $n=n_x+\chi_x n_p$ where $n_x$ ($n_p$) are densities of the excitons (polaritons) and $\chi_x$ is the excitonic weight in the polaritonic wavefunction. Since from the spectra we see that almost all the particles are from the LP-mode, we have $n\simeq\chi_x n_p$ \footnote{The assumption of no correlations used in Eq.\ref{Eq:Capacitor_Formula}, is not simply justified at low temperatures and at high densities. see Refs.  \cite{laikhtman_exciton_2009,shilo_particle_2013})}
\section{Conclusions}
In conclusion we showed that electrically gated WG-polaritons are highly tunable, where both the single polariton properties as well the dipolar polariton-polariton interaction strength can be easily controlled over a large range of values via local changes of the applied electric field. These polaritons are also fast propagating, inherently low-loss, and their in- and out-coupling can also be well controlled, making them highly suitable for both fundamental studies of interacting polariton systems and of interaction induced spatial correlations, as well as for future polariton-based logic circuitry. In particular, the dependence of the polariton energy on $\mathbf{F}$, and the ability to vary $\mathbf{F}$ locally due to the proximity of the polaritons to the surface, suggests that various potential and interaction landscapes can be tailored by choosing different layout for the top electrodes.     
\begin{acknowledgments}
We would like to acknowledge financial support from the U.S. Department of Energy: Office of Basic Energy Sciences - Division of Materials Sciences and Engineering, from the German- Israeli Foundation (GIF grant No. I-1277-303.10/2014), and from the Israeli Science Foundation (grant No. 1319/12). The work at Princeton University was funded by the Gordon and Betty Moore Foundation through the EPiQS initiative Grant GBMF4420, and by the National Science Foundation MRSEC Grant DMR-1420541.
\end{acknowledgments}
\nocite{treacy_dynamical_2002,schwarz_general_2012,andreani_exciton-polaritons_1994,burstein_confined_2012,semiconductor_2000,iotti_crossover_1997}
\appendix
\section{Experimental details}\label{App:exp_details}
The data which is presented in the paper was acquired during two different experiments. The difference between the experiments was in the temperature of the sample and in the method of optical excitation. The data which is presented in Fig.\ref{fig:fig3} of the paper was acquired under CW laser excitation at $\lambda=780nm$ and at 30 K. Here Only the TE-polarized PL (with respect to the out-coupler grating) was measured. The data which is presented in Fig.\ref{fig:fig4} was acquired at 5 K under pulsed excitation at $\lambda=775 nm$ where the pulse width and repetition rate were $300 ps$ and $200 kHz$ respectively and the PL was acquired during a $10 ns$ accumulation window starting at the rise of the laser pulse, using a gated intensified camera (PI-MAX). Here no polarizer was used during the acquisition of the PL and thus both the TE and the TM modes were measured.
The measurements were done using the k-space imaging setup which is illustrated in Fig.\ref{fig:A4}. The position of the excitation beam (spot size of $\sim 4$ $\mu m$ ) with respect to the grating is controlled by the mirrors M1 and M2. A pinhole located in the first image plane filters out photons which were emitted from regions around the grating. We used an x20 infinity corrected objective (NA=0.42), a Fourier lens of f=50 cm and an imaging lens of f=40 cm. 
\begin{figure}[h!]
  \centering
  \includegraphics[width=0.5\textwidth]{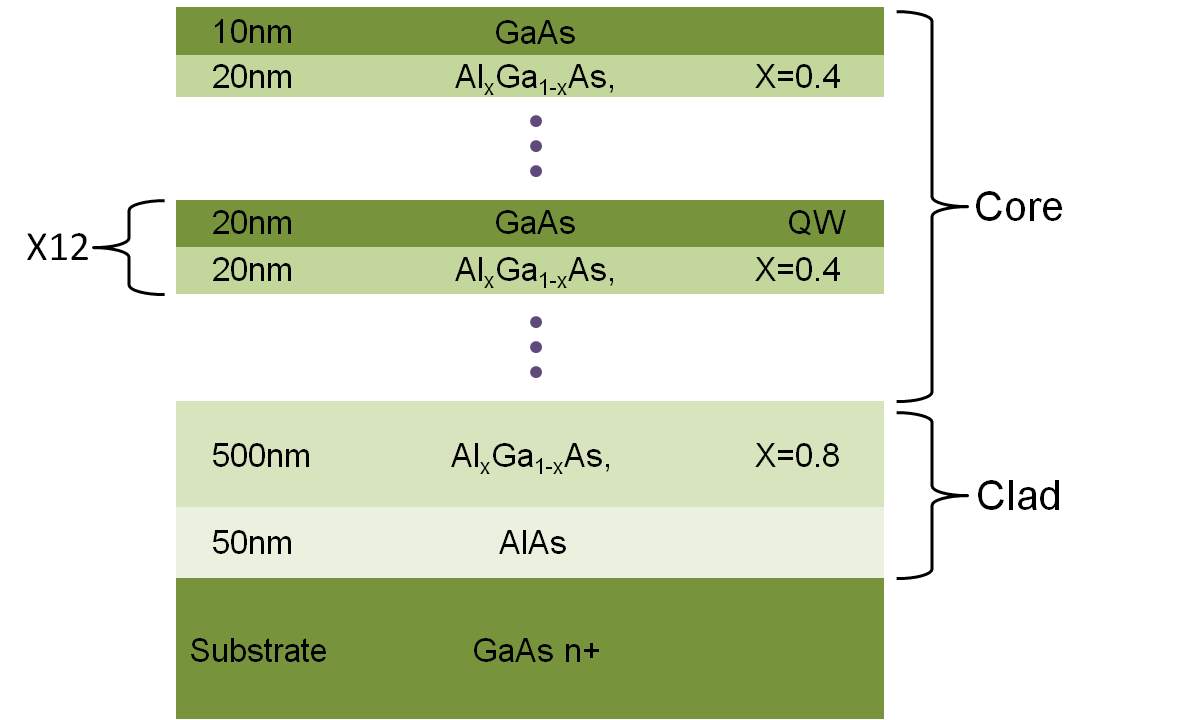}
  \caption{Layer structure the measured sample}
    \label{fig:A2}
\end{figure}
\begin{figure}[h!]
  \centering
  \includegraphics[width=0.5\textwidth]{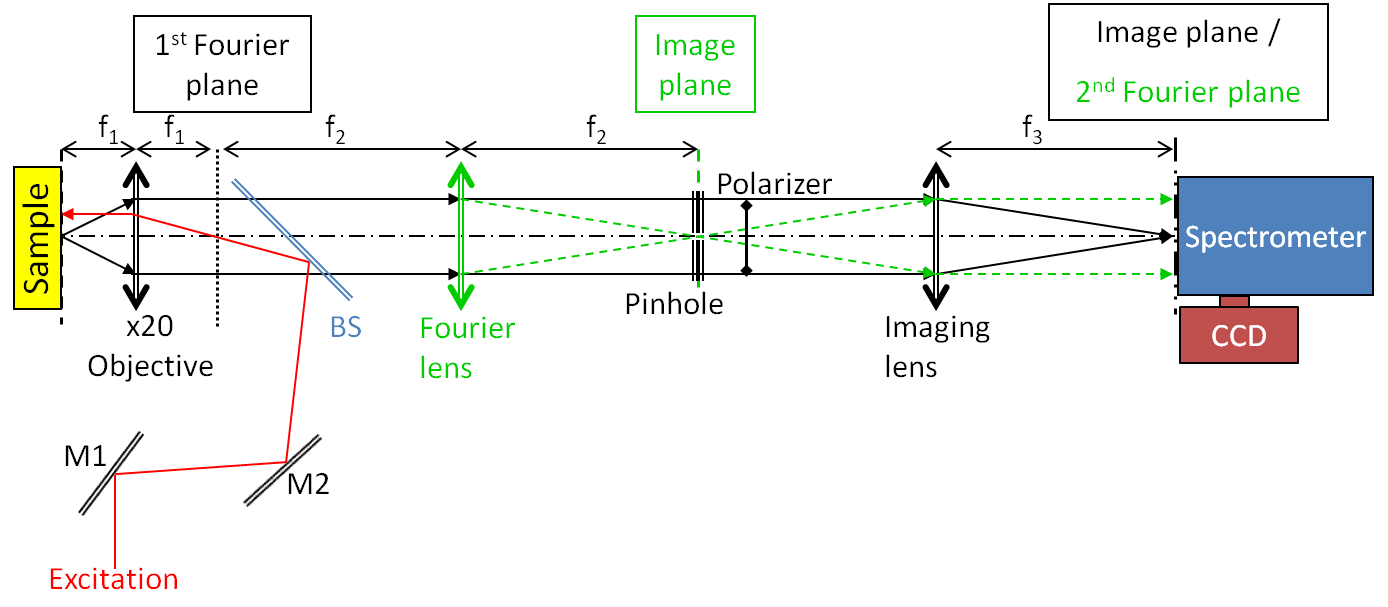}
  \caption{The Experimental setup. }
    \label{fig:A4}
\end{figure}
\section{Sample design and numerical simulations}\label{App:design_simulations}
\begin{figure}[h]
  \centering
  \includegraphics[width=0.5\textwidth]{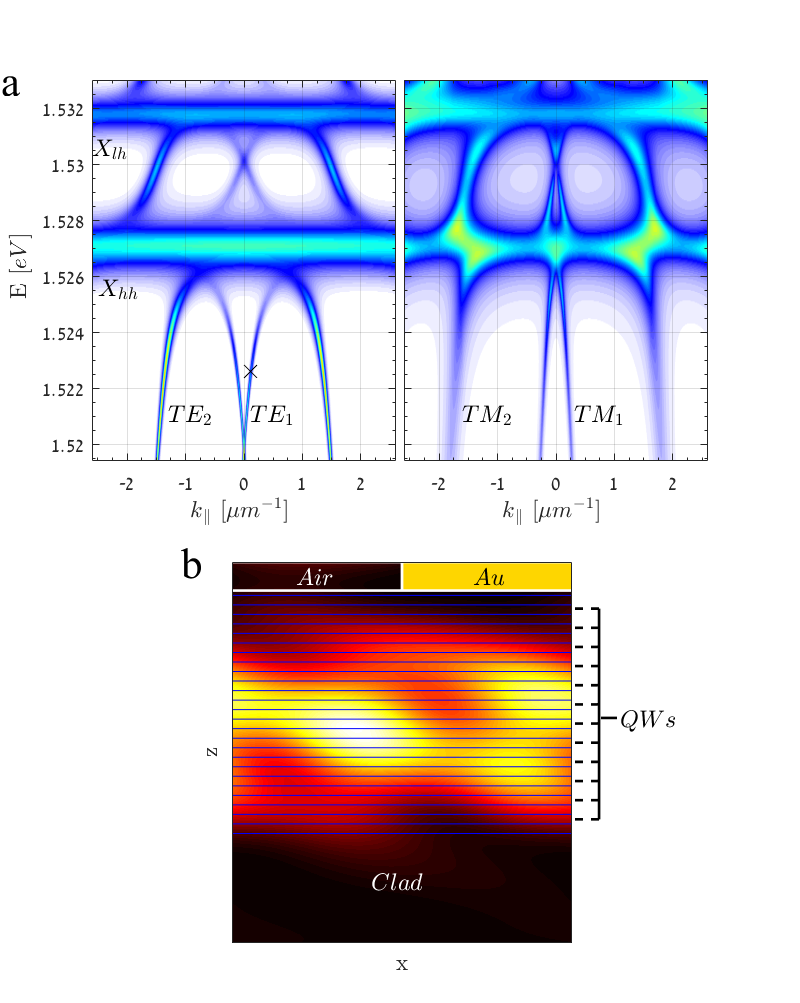}
  \caption{An RCWA Numerical calculations of our sample. a) Calculated dispersion for TE (left panel) and TM (right panel) polarizations. The $X_{hh}$ and $X_{lh}$ are marked as well as the 1st and 2nd WG modes at each polarization. The difference between the dispersions of the TE and the TM polaritons can be seen. b) The distribution of the electromagnetic field in one unit cell, calculated for the marked point at the top left panel.}
    \label{fig:A1}
\end{figure}

The design of our sample and some of the analysis were done with a numerical solver written in our lab. The solver, which calculates the response functions (reflection and transmission) of a stack of thin dielectric films with one periodic layer, is based on the method of rigorous coupled wave analysis (RCWA)\cite{treacy_dynamical_2002,schwarz_general_2012} and the transfer matrix formalism\cite{andreani_exciton-polaritons_1994}. In addition, the solver allows us to calculate the electromagnetic field distribution inside the WG from which the overlap between the photonic mode and each of the QWs can be deduced. In Fig.\ref{fig:A1} we present examples of typical calculations simulating the response of our sample. 
\section{TE and TM waveguide polaritons}\label{App:TE_TM}
Excitons  in GaAs QWs can be classified by the orientation of their optical dipole moment. T (L) excitons have an in-plane oriented optical dipole moment which is perpendicular (parallel) to their center-of-mass wave-vector, respectively. The optical dipole of Z-excitons is oriented along the growth direction of the QW. Due to symmetry considerations, Coupling of photons to Z-excitons is allowed only for the $X_{lh}$\cite{burstein_confined_2012}. TE-polaritons, which are waveguide polaritons consisting of a TE-polarized WG photonic mode,  can be composed of both T-$X_{hh}$ and T-$X_{lh}$, while TM-polaritons (with a TM-polarized WG photon), can be composed of L-$X_{hh}$, L-$X_{lh}$ and Z-$X_{lh}$. In addition, the interaction strength of the TM WG-mode and the L-$X_{hh}$ is proportional to the in-plane component of the electromagnetic polarization vector ($\propto \frac{k_z}{k}$)\cite{burstein_confined_2012}. In our system  $\frac{k_Z}{k}\sim\frac{1}{3}$  making $\Omega^{hh}_p(TE)=3\Omega^{hh}_p(TM)$, so the TE polaritons have a larger Rabi-splitting. This is the main reason we focus on those polaritons in this work. We also note that due to the reduced interaction between the TM-mode and the $X_{hh}$, the TE and TM polaritons have different dispersion, and are therefore easily separable (See Fig.\ref{fig:A1}). This is another feature which distinguishes WG-polaritons from MC-polaritons, which might be used to explore various types of inter-polarization parametric scatterings. 
\section{The coupled oscillator model}\label{App:coupled_osc}
The fitting to the model was as follows: from each measurement we extracted the bare mode energies $\omega_p(\beta)$, $\omega_{hh}(\beta,\mathbf{F})$ and $\omega_{lh}(\beta,\mathbf{F})$. The extraction of $\omega_p(\beta)$ was done by fitting the bare photon dispersion equation to the photon-like part of the LP dispersion. 
$\Omega_{hh}(0)$ and $\Omega_{lh}(0)$ are calculated from Ref. \cite{semiconductor_2000} to be $\hbar\Omega_{hh}(0)=\Gamma\cdot\sqrt{\frac{\hbar^2e^2}{8\epsilon L_w m_e}\frac{f_{hh}}{S}}=6.4$ $meV$ and $\hbar\Omega_{lh}(0)=3.5$ $meV$, where $L_W$ is the QW thickness, $m_e$ is the electron mass and $\epsilon$ is the dielectric constant. The oscillator strength per unit area of the heavy and light hole excitons were taken to be \cite{iotti_crossover_1997} $\frac{f_{hh}}{S}=6\cdot 10^{-4}$ and $\frac{f_{lh}}{S}=3\cdot 10^{-4}$ \AA $^{-2}$ respectively. The factor $\Gamma$ which was used as a fitting parameter represents the degree of overlap between the photonic mode and the QWs. After we found $\Gamma$, the spectra was fitted with $\Omega_{hh}(\bf{F})/\Omega_{hh}(0)$ and $\Omega_{lh}(\bf{F})/\Omega_{lh}(0)$ as the fitting parameters.
As the dependence of the coupled-modes splitting on the linewidth of the bare modes is proportional to the difference between the bare modes linewidths, and since these linewidths are roughly the same, the effect of the linewidth was neglected. 
\section{Calculation of the effective dipole length}\label{App:dipole}
Here we present the derivation of the effective dipole length which is presented as function of $\mathbf{F}$ in Fig.\ref{fig:fig1} of the paper and was used to extract the density in Fig.\ref{fig:fig4} \\
The interaction energy between two excitons can be described as follows:
\begin{equation}\label{Eq:A1}
    u(r)=\frac{e^2}{\epsilon}\sum_{\substack{x=e,h\\ y=e,h}}\epsilon_{xy}\int\frac{|\psi_1^x(z_1)|^2|\psi_2^y(z_2)|^2dz_1dz_2}{\sqrt{((z_1-z_2)^2+r^2)}}
  \end{equation}
where $z_i$ are the coordinates of the first ($i=1$ ) or second ($i=2$) interacting particle,  $\psi_i^x(z_i)$ are the envelope functions of the hole (x=h) or the electron ($x=e$) and $\epsilon_{xy} $ is given by\\
$\epsilon_{xy}=\begin{cases} 
      \;\;\:1 & x=y \\
     -1 & x\neq y \\
      \end{cases} $ \\
Under the mean field approximation the 
interaction energy is:
\begin{equation}\label{Eq:A2}
    E_{int}=n\int u(r) d^2r
\end{equation}
substituting Eq.\ref{Eq:A1} into Eq.\ref{Eq:A2} and integrating over $r$ we get:
\begin{equation}\label{Eq:A3}
  E_{int}=\frac{2\pi e^2 n}{\epsilon}\sum_{\substack{x=e,h\\ y=e,h}}\epsilon_{xy}\int|z_1-z_2| |\psi_1^x(z_1)|^2|\psi_2^y(z_2)|^2dz_1dz_2
\end{equation}
while in the process we neglected contributions from $r \to \infty$  due to the antisymmetric properties of $\epsilon_{xy}$.
Finally by comparing Eq.\ref{Eq:A3} to the capacitor formula of Eq.\ref{Eq:Capacitor_Formula}, we can identify the effective dipole length $d$ as:
\begin{equation}\label{Eq:A4}
    d=\sum_{\substack{x=e,h\\ y=e,h}}\epsilon_{xy}\int\frac{|z_1-z_2| |\psi_1^x(z_1)|^2|\psi_2^y(z_2)|^2 dz_1dz_2}{2}
\end{equation}
\begin{figure}[h!]
  \centering
  \includegraphics[width=0.5\textwidth]{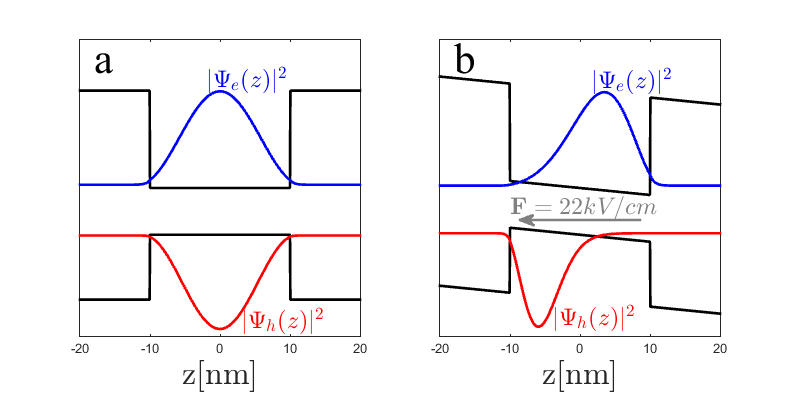}
  \caption{The probability distribution of the electron and the heavy hole at (a) $F=0$ and (b) $F=22 kV/cm$ calculated using our self consistent Schr\"odinger-Poisson solver.}
    \label{fig:A3}
\end{figure}
\section{PL of the Quantum Wells}\label{App:quality}
Here we present two PL spectra of the sample at the excitation point at the two different temperatures at which the measurements in the paper were taken. The light hole excitons can be seen in both, but are much more significant at the higher temperature of 30K
\begin{figure}[h]
  \centering
  \includegraphics[width=0.5\textwidth]{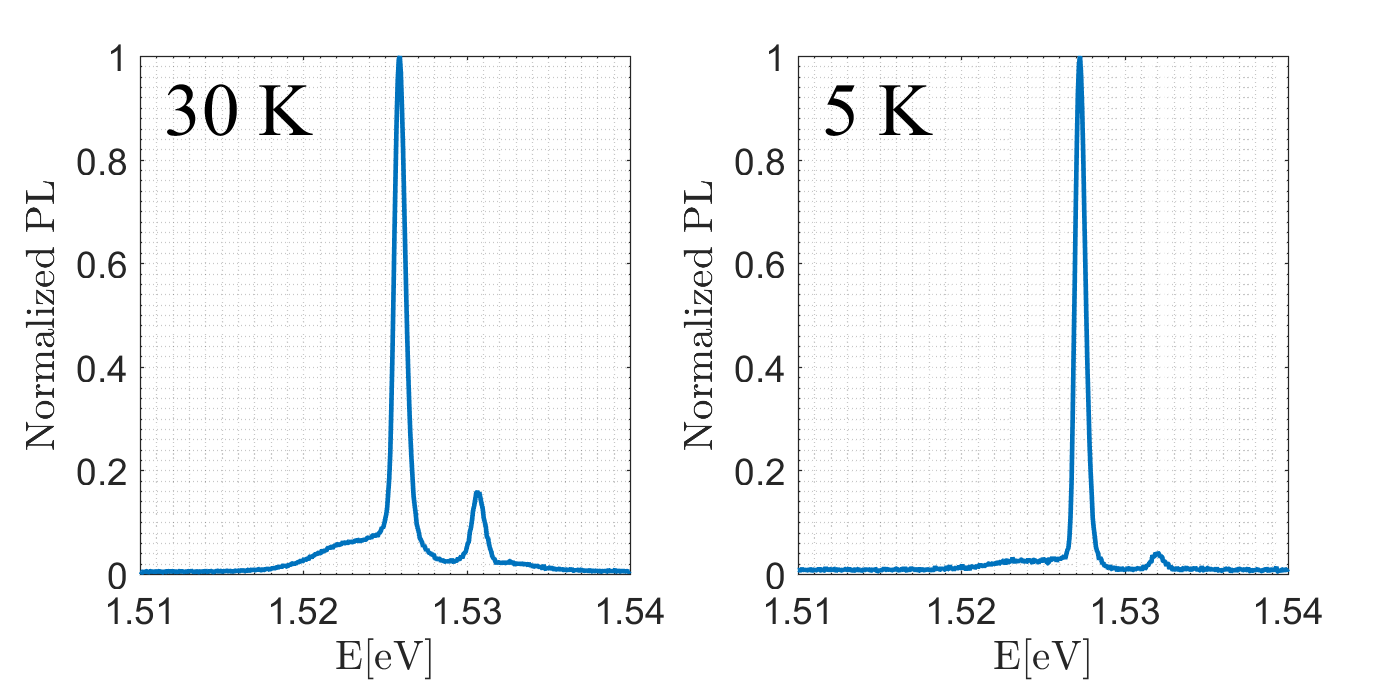}
  \caption{Normalized PL at 30 K (left) and 5 K (right)}
    \label{fig:A5}
\end{figure}
\FloatBarrier
\bibliographystyle{apsrev4-1}
\bibliography{oldbib}

\end{document}